# Ultrafast terahertz field control of the emergent magnetic and electronic interactions at oxide interfaces

A. M. Derrico[1,2], M. Basini[3], V. Unikandanunni[3], J. R. Paudel[1], M. Kareev[4], M. Terilli[4], T.-C. Wu[4], A. Alostaz[5], C. Klewe[6], P. Shafer[6], A. Gloskovskii[7], C. Schlueter[7], C. M. Schneider[5], J. Chakhalian[4], S. Bonetti[3,8,*], and A. X. Gray[1,*]

[1] *Department of Physics, Temple University, Philadelphia, Pennsylvania 19122, USA*

[2] *Department of Physics, University of California, Berkeley, California 94720, USA*

[3] *Department of Physics, Stockholm University, 10691 Stockholm, Sweden*

[4] *Department of Physics and Astronomy, Rutgers University, Piscataway, New Jersey 08854, USA*

[5] *Peter Grünberg Institut (PGI-6), Forschungszentrum Jülich GmbH, D-52425 Jülich, Germany*

[6] *Advanced Light Source, Lawrence Berkeley National Laboratory, Berkeley, California 94720, USA*

[7] *Deutsches Elektronen-Synchrotron, DESY, 22607 Hamburg, Germany*

[8] *Department of Molecular Sciences and Nanosystems, Ca' Foscari University of Venice, 30172 Venice, Italy*

*email: stefano.bonetti@unive.it, axgray@temple.edu

**ABSTRACT**

Ultrafast electric-field control of emergent electronic and magnetic states at oxide interfaces offers exciting prospects for the development of new generations of energy-efficient devices. Here, we demonstrate that the electronic structure and emergent ferromagnetic interfacial state in epitaxial $LaNiO_3/CaMnO_3$ superlattices can be effectively controlled using intense single-cycle THz electric-field pulses. We employ a combination of polarization-dependent X-ray absorption spectroscopy with magnetic circular dichroism and X-ray resonant magnetic reflectivity to measure a detailed magneto-optical profile and thickness of the ferromagnetic interfacial layer. Then, we use time-resolved and temperature-dependent magneto-optical Kerr effect, along with transient optical reflectivity and transmissivity measurements, to disentangle multiple correlated electronic and magnetic processes driven by ultrafast high-field (~1 MV/cm) THz pulses. These processes include an initial sub-picosecond electronic response, consistent with non-equilibrium Joule heating; a rapid (~270 fs) demagnetization of the ferromagnetic interfacial layer, driven by THz-field-induced nonequilibrium spin-polarized currents; and subsequent multi-picosecond dynamics, possibly indicative of a change in the magnetic state of the superlattice due to the transfer of spin angular momentum to the lattice. Our findings shed light on the intricate interplay of electronic and magnetic phenomena in this strongly correlated material system, suggesting a promising avenue for efficient control of two-dimensional ferromagnetic states at oxide interfaces using ultrafast electric-field pulses.

## I. INTRODUCTION

The study and design of complex-oxide heterostructures that exhibit emergent electronic and magnetic phenomena at their interfaces has become an active area of research in condensed matter physics and materials science [1-7]. Particularly intriguing are material systems where the intricate interplay among different degrees of freedom at the interface leads to unique functional properties and states [8-14]. Ultrafast electric-field control of these emergent electronic and magnetic states offers the potential for developing new types of atomically-thin spintronic devices that support transient states, require minimal energy consumption, and operate at speed limits governed only by the fundamental laws of physics [15-18].

The burgeoning field of ultrafast terahertz (THz) science has been exploring this promising avenue using both resonant and non-resonant THz electric-field pulses to understand and control the ultrafast processes and dynamics of charge carriers, spins, and phonons in a wide variety of material systems [18-38]. Notably, several recent studies have showcased the effectiveness of THz excitation in triggering and controlling seminal physical phenomena in oxides, including insulator-metal transition [18,19], ferroelectricity [20], and transient superconductivity [21,22]. However, investigations of THz interactions with magnetically ordered materials have mainly focused on the dynamics and coherent spin control in bulk single-crystal, bulk-like thin-film insulating antiferromagnets [23-33], and metallic ferromagnets [34-38], but not in epitaxially engineered interfaces hosting strong correlations between electronic and magnetic degrees of freedom.

Here, we consider ultrafast THz control of the emergent ferromagnetic order induced by interface engineering in an archetypal strongly-correlated heterostructured oxide superlattice consisting of antiferromagnetic $CaMnO_3$ and paramagnetic $LaNiO_3$ [39-43]. The use of intense single-cycle THz electric-field pulses is designed to effectively simulate the action of an ultrafast picosecond current pulse, which has been recently shown to induce magnetization switching in a metallic ferromagnetic heterostructure [44]. We demonstrate that the electronic structure and emergent interfacial ferromagnetic state at the interface can be effectively controlled using such ultrafast THz electric-field pulses.

For our study, we utilize a combination of polarization-dependent X-ray absorption spectroscopy with magnetic circular dichroism (XAS/XMCD) and soft X-ray resonant magnetic reflectivity (XRMR) [45] to measure the detailed magneto-optical profile of the $LaNiO_3/CaMnO_3$ interface and determine the characteristic thickness of the ferromagnetic interfacial layer. We then employ a combination of temperature dependent time-resolved magneto-optical Kerr effect (tr-MOKE), optical reflectivity, and transmissivity measurements to disentangle multiple interrelated electronic and magnetic processes driven by ultrafast high-field (~1 MV/cm) THz electric-field pulses and evolving with different characteristic timescales.

Our measurements reveal several distinct ultrafast responses to THz excitation. During the first picosecond, an electronic response marked by non-equilibrium Joule heating due to electron-electron scattering occurs simultaneously with the THz pulse [46]. It is accompanied by a sudden partial metallization of the superlattice via direct interband tunnelling [18,47,48]. At temperatures below $T_C$ (~80 K), these sub-ps electronic dynamics become strongly modulated by a concomitant magnetic response indicative of rapid (~270 fs) demagnetization of the ferromagnetic interfacial layer driven by THz-field-induced nonequilibrium spin-polarized currents. Subsequently, the system experiences electron-phonon thermalization. Thus, in addition to the initial sub-picosecond response, slower, highly temperature-dependent magneto-optical dynamics emerge on a multi-picosecond timescale, possibly signifying a change in the magnetic state of the superlattice due to the transfer of spin angular momentum to the lattice.

Our findings show that the multiple interrelated non-equilibrium electronic and magnetic processes originating at the LaNiO$_3$/CaMnO$_3$ interface can be disentangled in the time domain. Additionally, a correlation between the charge and spin dynamics exists in this material system, and the electronic effects observed below T$_C$ (~80 K) are strongly modulated by the interfacial magnetism. This connection suggests a pathway for efficiently switching the two-dimensional ferromagnetic states at oxide interfaces using ultrafast THz electric fields.

## II. RESULTS
### a. Element-specific and depth-resolved profiling of the interfacial ferromagnetism

In epitaxial LaNiO$_3$/CaMnO$_3$ superlattices, strongly correlated physics and competing ferromagnetic exchange interactions are intertwined at the interface, giving rise to quasi-2D ferromagnetism [39]. The emergent ferromagnetic order arises due to the interfacial charge transfer between LaNiO$_3$ and CaMnO$_3$ [43] and can be tuned by modulating the conductivity of LaNiO$_3$, which undergoes a metal-insulator transition in the ultrathin (few-unit-cell) limit [49]. Thus, in superlattices with an above-critical LaNiO$_3$ thickness (d ≥ ~4 u.c.), ferromagnetism mediated by double-exchange interaction is stabilized on the interfacial Mn sites [39]. Conversely, in the superlattices with a below-critical LaNiO$_3$ thickness (d < ~4 u.c.) the charge transfer is suppressed and, therefore, only a very weak residual ferromagnetic signal is detected in some samples [40] due to defect-mediated phenomena.

For this study, a LaNiO$_3$/CaMnO$_3$ superlattice with ten repetitions containing bulk-like metallic LaNiO$_3$ layers with a thickness of eight unit cells (8 u.c.) and 4 u.c. of insulating CaMnO$_3$ was synthesized on a single-crystalline LaAlO$_3$ (001) substrate using pulsed laser interval deposition [50]. Epitaxial growth was monitored in situ via reflection high-energy electron diffraction (RHEED). Details of the ex-situ characterization of the superlattice using X-ray diffraction (XRD), resonant and non-resonant soft X-ray reflectivity (SXR), synchrotron-based hard X-ray photoelectron spectroscopy (HAXPES) [51], as well as electronic transport and SQUID magnetometry (M vs. T), are presented in the Methods section and the Supplementary Information (Figures S1-S4). An additional control sample comprised the same number of repetitions (10) and the same CaMnO$_3$ layer thickness (4 u.c.), but with ultrathin (2 u.c.-thick) insulating LaNiO$_3$ layers, was synthesized in the same batch. For brevity, we will refer to these two superlattices as 8LNO/4CMO and 2LNO/4CMO, respectively, indicating the LaNiO$_3$ thickness.

In order to unequivocally establish the interfacial origin of the ferromagnetic signal detected in the 8LNO/4CMO superlattice by SQUID magnetometry (see Figure S4), we carried out element-specific and depth-resolved magnetic characterization and profiling of the sample using XAS/XMCD and XRMR at the high-resolution (100 meV) Magnetic Spectroscopy beamline 4.0.2 at the Advanced Light Source [52].

The XAS spectrum of the Mn $L_{2,3}$ absorption edges shown in the top panel of Figure 1a exhibits excellent agreement with prior studies [42,43]. The spectrum reveals a low-energy (640.5 eV) spectral feature attributed to the presence of Mn$^{3+}$ cations, suggesting a mixed (3+/4+) Mn valence state in CaMnO$_3$. This mixed valence state is required for the establishment of the ferromagnetic double exchange interaction [39-43].

The resulting ferromagnetic state of Mn is then observed in the XMCD spectrum shown in the bottom panel. Due to the depth-averaging nature of the XAS/XMCD technique, the XMCD signal originating from the ultrathin buried interfacial regions is weak (~0.1%) compared to that of typical manganite films showing mostly depth-uniform magnetization [53]. However, this highly localized magnetic signal can be amplified by carrying out depth-resolved $q_z$-dependent

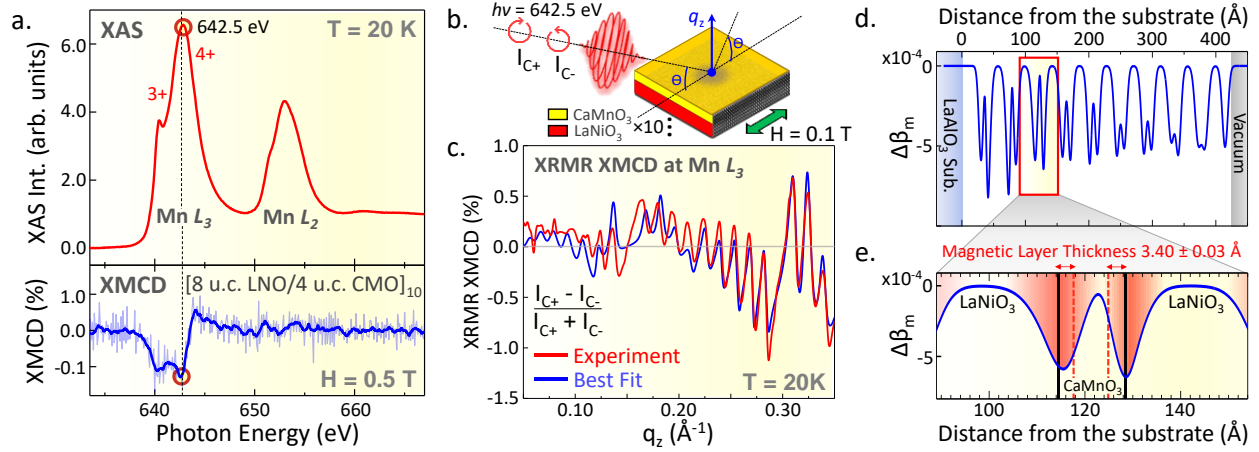

**Figure 1 | Element-specific and depth-resolved profiling of the interfacial ferromagnetism. a.** Top panel: Bulk-sensitive Mn $L_{2,3}$-edge XAS spectrum, measured in luminescence detection mode at T = 20 K and probing the entire depth of the 8LNO/4CMO superlattice, reveals a mixed (3+/4+) Mn valence state in the CaMnO$_3$ layers. Bottom panel: XMCD spectrum measured in an in-plane magnetic field of 0.5 T exhibits a significant magnetic signal of up to −0.15% at the Mn $L_3$ edge (642.5 eV). The light-blue spectrum represents raw data while the solid blue curve has been smoothed using the Savitzky-Golay method. **b.** Schematic diagram of the XRMR-XMCD measurements in the specular reflection geometry. **c.** XRMR-XMCD asymmetry and the best fits to the experimental data measured at the resonant photon energy of the Mn $L_3$ edge. Self-consistent fitting of the data yields a detailed magneto-optical profile of the superlattice, as shown in panels d and e. **d.** Depth-resolved magneto-optical profile of the entire superlattice given by the modulation of the magnetic dichroism of the x-ray optical constant $\Delta\beta_m$. **e.** Detailed magneto-optical profile of the CaMnO$_3$ layer and two adjacent LaNiO$_3$ layers in the near-central region of the superlattice. Typical thickness of the interfacial ferromagnetic layer is 3.40±0.03 Å with a 2.39±1.49 Å Névot-Croce-type interdiffusion present on either side.

XRMR spectroscopy, which, in conjunction with X-ray optical modeling, can be used to derive the detailed magneto-optical profile of the superlattice [45,54].

The above-mentioned XRMR measurements were carried out in the specular reflection geometry depicted schematically in Figure 1b. A resonant photon energy of 642.5 eV, corresponding to the maximum of the XMCD signal at the Mn $L_3$ absorption edge, was used. The percent magnetic asymmetry ($I_{C+} - I_{C-} / I_{C+} + I_{C-}$) was recorded as a function of momentum transfer $q_z$. The measurements were carried out in an applied in-plane magnetic field of 0.1 T and at a sample temperature of 20 K, which is well below the reported T$_C$ for this system (~80 K). An additional XRMR spectrum at the photon energy corresponding to non-resonant excitation (620 eV) was also recorded and analyzed (see Fig. S2 in the Supplementary Information).

The resulting experimental XRMR-XMCD vs. $q_z$ spectrum is depicted as a red curve in Figure 1c. It is crucial to note that the measured $q_z$ range spans both the first- and second-order Bragg conditions (at ~0.15 and ~0.30 Å$^{-1}$, respectively). Consequently, it encompasses detailed depth-resolved information concerning the interfacial magnetic structure of the sample [54,55].

The XRMR-XMCD spectrum shown, along with the non-resonant and resonant X-ray reflectivity spectra in Figure S2 of the Supplementary Information, were self-consistently fitted using the x-ray reflectivity analysis program ReMagX [54]. The best theoretical fit is represented by the blue-colored spectrum in Figure 1c, demonstrating a good agreement with the experimental data in terms of feature amplitudes, relative phases, and lineshapes. It is worth noting that the utilization of $q_z$-dependent $I_{C+} - I_{C-} / I_{C+} + I_{C-}$ spectra significantly enhances the sensitivity of the fitting process. These spectra often reveal intricate lineshapes and numerous sharp modulations with varying amplitudes and shapes across the entire $q_z$ range, thereby severely constraining the fitting model and facilitating sub-Ångstrom-level depth resolution [55].

A self-consistent X-ray magneto-optical profile of the superlattice resulting from the fitting of three $q_z$-dependent XRMR spectra (non-resonant, resonant, and XRMR-XMCD) is depicted in Figure 1d. The profile represents the depth-dependent variation (x-axis) of the X-ray optical constant $\Delta\beta_m$, which quantifies the magnitude of the modulation of the magnetic dichroism of the x-ray absorption coefficient $\beta$ at the resonant photon energy of the Mn $L_3$ edge (642.5 eV). The all-negative values are consistent with the traditional sign convention for representing XMCD signal at the $L_3$ edge.

The immediate vicinity of the LaAlO$_3$ substrate (at the distances from the substrate of approximately 0-28.5 Å, as defined in Figure 1d) is occupied by the first 8 u.c.-thick paramagnetic LaNiO$_3$ layer. This layer exhibits no XMCD ($\Delta\beta_m$) signal, except for a narrow (~2.39 Å) interdiffusion region where it interfaces with the first CaMnO$_3$ layer. In the subsequent depth range (~28.5-42.7 Å), we find the aforementioned 4 u.c.-thick CaMnO$_3$ layer, which exhibits a distinctive double-peaked excursion in $\Delta\beta_m$. Such a magneto-optical profile indicates a significant and highly localized increase in the net magnetic moment of Mn within the interfacial regions of the CaMnO$_3$ layer. This LaNiO$_3$/CaMnO$_3$ bilayer profile is then replicated ten times, consistent with the nominal layering of the superlattice, with the exception of the uppermost CaMnO$_3$ layer where only the bottom interface is ferromagnetic.

In order to better elucidate the local magnetic structure, in Figure 1e we plot a more detailed magneto-optical profile of the CaMnO$_3$ layer (as well as the two adjacent LaNiO$_3$ layers) in the near-central region of the superlattice. The figure illustrates that the maxima of the magnetic signal are located within the interfacial unit cells of CaMnO$_3$ on both "ends" of the layer. The thickness of the ferromagnetic region in CaMnO$_3$, as defined by the fitting model, is 3.40±0.03 Å, which corresponds to approximately one pseudo-cubic unit cell [42,43,56]. However, there is a significant broadening of the $\Delta\beta_m$ peak, characterized by the Névot-Croce-type interdiffusion [57], with a characteristic width of 2.39±1.49 Å present on both sides of the ferromagnetic region. Therefore, although the ferromagnetism clearly originates in the interfacial unit cell of CaMnO$_3$, the total extent of the ferromagnetic signal in a real-life (non-ideal) sample is on the order of 2-3 u.c.

It is important to highlight that the minor asymmetries in the widths and amplitudes of the magnetic signal observed for the top and bottom interfaces in Figures 1d-e are typical for such superlattices. These differences have been previously ascribed to structural asymmetries, roughness, and interface reconstruction [55,58,59]. It is also worth noting that a gradual degradation in the quality of the interfaces is observed in Figure 1d for several of the upper bilayers in the superlattice. It is characterized by the reduction of the maximum magnitudes of $\Delta\beta_m$ observed at the interfaces, as well as by the gradual merging of the two initially sharp interfacial peaks in $\Delta\beta_m$. This degradation is likely due to the cumulative increase in the number of defects, such as elemental interdiffusion and interface roughness/steps, which is typical for the PLD growth of superlattices consisting of many tens and/or hundreds of atomic layers [60,61]. Nevertheless, it is important to note that this effect becomes prominent only in the top 2-3 bilayers and, even so, the observed ferromagnetic signal in the affected CaMnO$_3$ layers still clearly originates from the interfacial regions.

In summary, we used a combination of XAS/XMCD spectroscopy and XRMR to measure the detailed magneto-optical profile of the conducting 8LNO/4CMO superlattice and to show that the ferromagnetism in this system originates and resides mostly in the interfacial unit cells of $CaMnO_3$. In the following section of this work, we demonstrate that this emergent interfacial phenomenon can be controlled using ultrafast high-field (~1 MV/cm) THz E-field pulses and discuss multiple interrelated electronic and magnetic processes driven by the excitation.

### b. Ultrafast THz dynamics and control of the interfacial ferromagnetism

To induce and examine the ultrafast E-field driven electronic and magnetic dynamics in the $LaNiO_3$/$CaMnO_3$ superlattices, we used single-cycle THz pump pulses generated in an OH1 organic crystal through the optical rectification of 1300 nm radiation emitted by an optical parametric amplifier [62]. Figure 2a presents the resulting THz pulse, which exhibited a peak E-field reaching 1 MV/cm and a central frequency of approximately 1.9 THz (see inset). The measurement and calibration of the pulse were conducted using standard electro-optic sampling techniques [63].

A portion of the 800-nm beam produced by the same Ti:sapphire amplifier served as a probe for investigating the electronic and magnetic dynamics. Three detection schemes, implemented in the same experimental setup and illustrated schematically in Figures 2a-e, were employed. The ultrafast reflectivity (at 45⁰ incidence) and transmissivity (at normal incidence) measurements were utilized to examine the electronic response of the sample, while the time-resolved magneto-optical Kerr effect (tr-MOKE) technique (at 45⁰ incidence) was employed to probe the magnetic response at the interface. The measurements were conducted within an in-plane applied magnetic field of approximately 0.47 T and in a closed cycle cryostat, enabling temperature control from 20 to 300 K.

Figures 2b and 2c depict time-delay traces of the percent change in the near-IR ($\lambda = 800$ nm) reflectivity ($\Delta R/R$) and transmissivity ($\Delta T/T$), respectively, recorded at room temperature (300 K) on the conducting 8LNO/4CMO superlattice. These delay traces represent the predominantly electronic (non-magnetic) response of the system to the THz excitation, due to the above-$T_C$ temperature of the sample and the nominally magnetically insensitive detection modes (R and T).

The initiation of the electronic response is observed simultaneously with the onset of the THz excitation pulse shown in Figure 2a. As the data in Figures 2b and 2c demonstrate, at room temperature, the observed dynamics are dominated by processes occurring within the temporal window when the THz pulse is present in the sample (~1 ps). Specifically, we observe a substantial increase in the relative optical reflectivity ($\Delta R/R$) that is mirrored by a concomitant decrease in the relative transmissivity ($\Delta T/T$). This suggests an electronic response consistent with non-equilibrium Joule heating due to electron-electron scattering [46] as well as an E-field-driven modification of the electronic structure leading to a transient reduction of the near-IR absorption, analogous to a sudden metallization of the sample via direct electronic interband tunnelling [18,47,48]. These effects are observed simultaneously with the THz excitation pulse and diminish considerably once the THz pulse is no longer active within the sample (past t = 1 ps), giving way to a stable, long-lived transient state with a partially collapsed bandgap (2 ps < t < 12 ps). This long-lived partially-metallic state, characterized by increased R and decreased T, persists for picoseconds after the THz pulse. This suggests a slow relaxation of the elevated lattice temperature associated with the partially-metallic state [64]. Similar dynamics have been observed recently in another strongly-correlated oxide system, $VO_2$, where THz pulses were shown to induce an insulator-metal transition assisted by similar purely-electronic mechanisms [18].

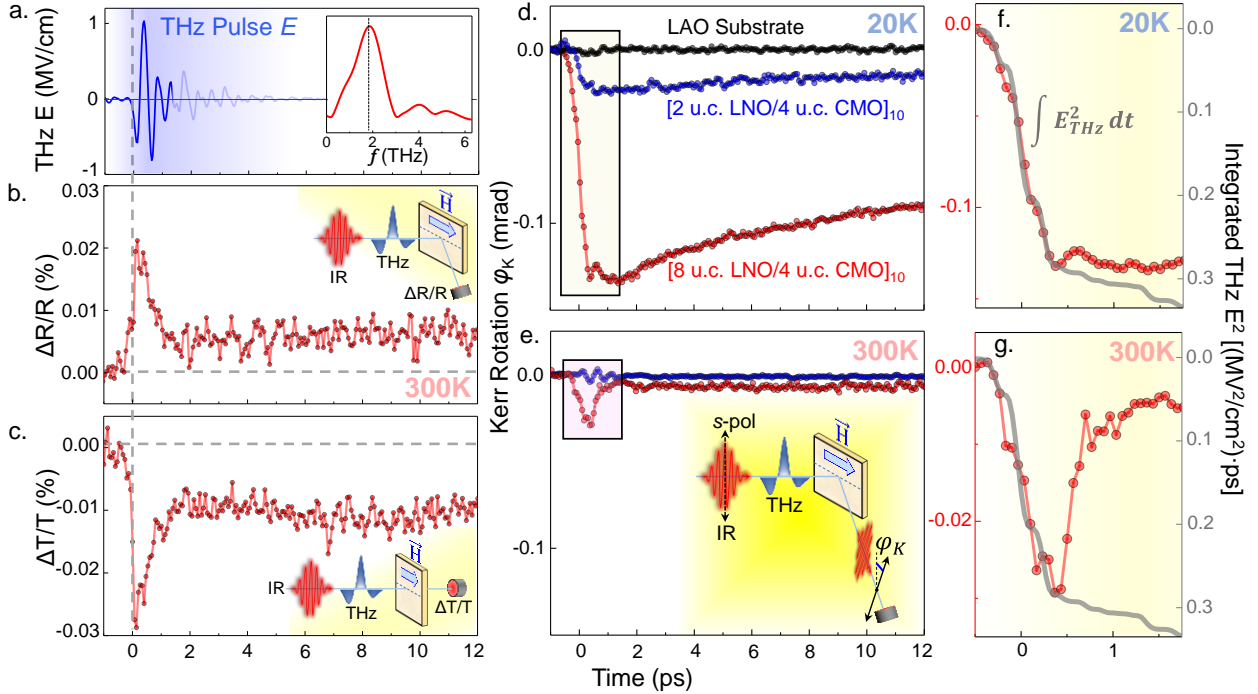

**Figure 2 | Electronic and magnetic dynamics in the LaNiO$_3$/CaMnO$_3$ superlattices induced by an ultrafast THz $E$-field pulse. a.** The THz pulse $E$-field waveform, as measured and calibrated via standard electro-optic sampling (EOS). Shaded (light blue) features after ~1.5 ps are artefacts due to multiple reflections in the EOS crystal. The Fourier-transformed pulse is shown in the inset. **b.** and **c.** Ultrafast THz-pump near-IR-probe (800 nm) time-delay traces of the relative percent changes in reflectivity (b) and transmissivity (c) showing an ultrafast transient response followed by a long-lived transient state characteristic of a sudden metallization of the superlattice via direct interband tunnelling. Experimental geometries are shown in the insets. **d.** and **e.** Time-delay traces of the Kerr rotation $\phi_K$, as measured via tr-MOKE for the two superlattice samples and the bare LaAlO$_3$ substrate, below (d) and above (e) the Curie temperature ($T_C$). Experimental geometry is shown in the inset (e). In addition to the expected temperature dependence of the magnetic response, the observed dynamics show strong LaNiO$_3$ thickness dependence. The LaAlO$_3$ substrate shows almost no measurable response to the THz pulse. **f.** and **g.** The initial sub-ps response of the delay traces follows in shape the integral of the square of the THz pulse waveform and is present in both below-$T_C$ (f) and above-$T_C$ (g) dynamics.

Figures 2d and 2e depict the dynamic magnetic responses (Kerr Rotation $\phi_K$) of the 8LNO/4CMO and 2LNO/4CMO superlattices at two temperatures: 20 K (below $T_C$) and 300 K (above $T_C$), respectively. At first glance, it is immediately evident that the overall amplitudes of these responses are strongly temperature- and LaNiO$_3$-thickness dependent.

Specifically, the strongest magnetic response to the THz excitation is observed for the 8LNO/4CMO superlattice containing metallic LaNiO$_3$ layers (red curve in Fig. 2d), where interfacial ferromagnetic order is observed via XRMR-XMCD at T = 20 K (see Fig. 1). A much weaker (by ~80%) magnetic response is observed for the 2LNO/4CMO superlattice containing insulating LaNiO$_3$ layers (blue curve), where only a very weak residual ferromagnetic signal due to defect-mediated phenomena is expected.

The ~80% reduction in the signal is quantitatively consistent with the static SQUID magnetometry measurements on similar 8LNO/4CMO and 2LNO/4CMO samples by Flint *et al.* [40] as well as XMCD measurements by Paudel *et al.* [43]. Such quantitative agreement with the prior studies of interfacial ferromagnetism in the LaNiO$_3$/CaMnO$_3$ superlattices is the first indication that the observed THz-driven magnetic dynamics originate at the interface.

The second piece of evidence supporting the interfacial origin is observed in the temperature dependence of the overall amplitudes of the dynamic magnetic response, when comparing Figures 2d (below-T$_C$) and 2e (above-T$_C$). Namely, the tr-MOKE effect is most pronounced and exhibits a rich multistep evolution for the 8LNO/4CMO superlattice in its ferromagnetic state at T = 20 K. Conversely, it is significantly weaker at T = 300 K, where it essentially mimics the purely-electronic response, as seen in ΔR/R and ΔT/T in Figures 2b-c.

It is important to note that the initial sub-ps response observed during the first 1 ps of the delay traces follows in shape the integral of the square of the THz pulse waveform and is present in both below-T$_C$ (Fig. 2f) and above-T$_C$ (Fig. 2g) dynamics. This further confirms the purely-electronic origin of this dynamic component of the response, but also suggests that the sub-ps magnetic response to the THz excitation is modulated by this THz-driven electronic effect.

The underlying reason for the observed correlated behavior of the electronic and magnetic dynamics is the fact that the nonequilibrium current induced at the ferromagnetic interface by the THz pulse is, in fact, spin-polarized [65] and, therefore, becomes detectable not only in the R and T, but also in the Kerr Rotation ($\phi_K$) channel [66]. To further support this, our fluence-dependent measurements shown in Figure S5 of the Supplementary Information demonstrate that the amplitude of the response scales with the square of the THz peak E-field, which is an expected functional dependence for energy dissipation due to scattering processes within a THz-driven spin current [66].

The final key observation from Figure 2d is that, at the temperatures below T$_C$, an additional dynamic component characterized by a slower fall time (~1.1 ps) and consequent exponential recovery arises in the tr-MOKE signal. This component is most evident in the red time-trace (8LNO/4CMO) in Figure 2d. In order to investigate its temperature dependence and ascertain its origin, we carried out systematic temperature-dependent tr-MOKE and near-IR transient reflectivity measurements, which are shown and analyzed in Figures 3 and 4 below.

Figure 3a shows the time-resolved magneto-optical response of the sample measured via tr-MOKE at seven temperatures varying from 300 K (room temperature) to 20 K. The initial sub-ps response to the THz excitation observed during the first 1 ps of the delay traces is observed for all temperatures. For brevity, from now on, we will refer to this dynamic component as the "fast" peak. In addition to this feature, the 300 K delay trace (black curve), which essentially mimics the purely-electronic response as observed in ΔR/R and ΔT/T, also exhibits the long-lived transient state characterized by the stable plateau in the Kerr rotation ($\phi_K$) from ~1 to 5 ps.

Due to its simplicity, the 300 K delay trace can be effectively fitted using two functional components – a Gaussian lineshape representing the *fast* peak (red curve in Figure 3b) and a quasiconstant component characterized by a step function broadened by the width of the THz pulse, ~1 ps (shown as the orange curve). The total fit to the data is shown as a solid gray curve in Figure 3b. Although this fitting model does not consider the fine (sub-ps) structure of the THz pulse (see Fig. 1a), it effectively accounts for the major features of the delay trace corresponding to different THz-induced physical phenomena in the sample. This model was recently used to describe the ultrafast photoinduced insulator-metal transition in SmNiO$_3$ [67].

As temperatures decrease towards T$_C$ = 80 K, an additional dynamic component characterized by a slower fall time and consequent exponential recovery emerges, becoming pronounced precisely at T$_C$. Incorporating this component into our fitting model is essential for accurately describing the shape of the 80 K delay trace. It is shown as the blue curve in Figure 3c. This adjustment is also necessary for fitting all lower-temperature delay traces, as illustrated by the 20 K data in Figure 3d. Henceforth, we will refer to it as the "slow" component that arises and grows at the onset of interfacial ferromagnetism in the $LaNiO_3$/$CaMnO_3$ superlattice.

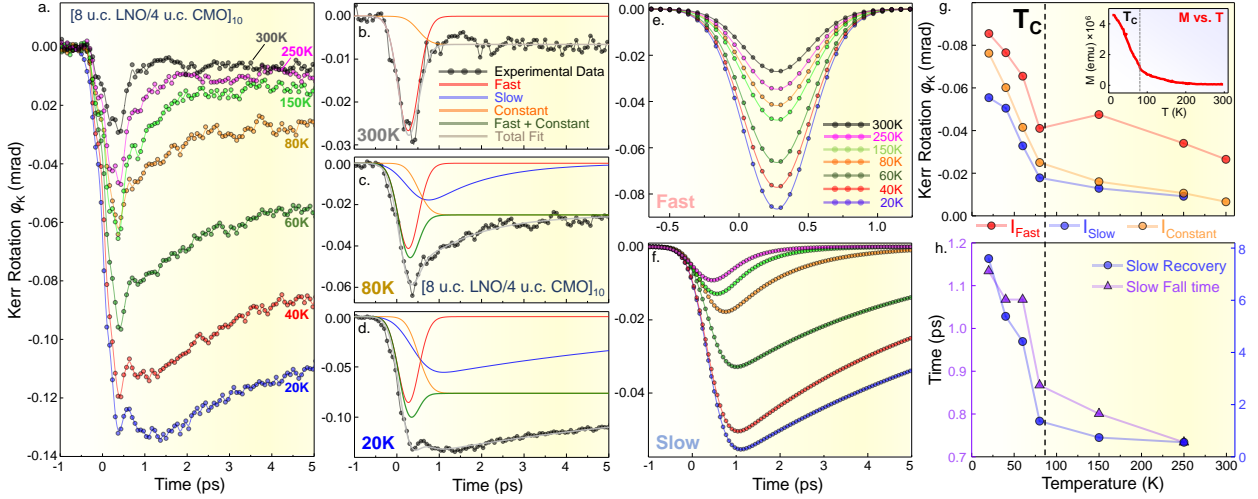

**Figure 3 | Temperature-dependent tr-MOKE and the origins of multiple dynamic magnetic responses. a.** Temperature dependent tr-MOKE delay traces recorded at seven temperatures varying from 300 to 20 K. **b.** Spectral decomposition of the room-temperature (300 K) delay trace into the initial *fast* and quasiconstant components. **c.** Spectral decomposition of the delay trace recorded at $T_C$ = 80 K requires an additional dynamic component characterized by a slower fall time and consequent exponential recovery (blue curve). **d.** Temporal decomposition of the 20 K delay trace requires all three dynamic components. **e.** All the temperature-dependent fits of the *fast* dynamic component, exhibiting uniform temporal behavior. **f.** Temperature-dependent fits of the *slow* dynamic component, which exhibits onset at 250 K and grows at the onset of interfacial ferromagnetism (T < 80 K). **g.** and **h.** Amplitudes (g) and the characteristic time constants (h) of the dynamic components of the tr-MOKE delay traces plotted as functions of temperature. All parameters exhibit a sharp decline and a change in slope at 80 K, marking the critical temperature ($T_C$) for the interfacial magnetic order, consistent with static SQUID magnetometry measurements shown in the inset (reproduced with permission from Ref. 68).

Figures 3e and 3f depict all the temperature-dependent fits of the *fast* and *slow* components, respectively. For the *fast* peak, the Gaussian width and the center position (270 fs) characterizing the fall time of the effect are independent of temperature. This is expected because of the electronic nature of the underlying physical effect, which involves non-equilibrium Joule heating and THz-field-induced spin-polarized currents leading to the ultrafast demagnetization of the ferromagnetic interfacial layer. Consequently, the temporal profile of this dynamic component is exclusively determined by the shape (width and time-zero) of the THz pulse.

On the other hand, the *amplitude* of the fast peak displays a pronounced temperature dependence, as depicted in Figure 3g (red symbols). It is readily apparent that this behavior aligns with the static SQUID magnetometry (M vs. T) measurements, revealing a sharp decline and a change in slope at 80 K, marking the critical temperature ($T_C$) for the interfacial magnetic order (see inset in Figure 3g as well as Figure S4 in the Supplementary Information). The time-resolved data also accurately captures the presence of some residual magnetic signal observed between $T_C$ and 200-250 K, which has been documented in prior studies and attributed to the presence of multiple competing magnetic interactions in the superlattice [68]. Thus, the agreement between static magnetometry and dynamic tr-MOKE data provides additional compelling evidence that the THz pulse directly influences the interfacial ferromagnetic order on the ultrafast scale.

The amplitudes of the *slow* and the quasiconstant components similarly exhibit temperature dependence that is consistent with the magnetometry (M vs. T) measurements, showing an abrupt change in slope at $T_C = 80$ K and some residual signal up to 200-250 K (blue and orange symbols in Fig. 3g). Importantly, the *slow* component also exhibits qualitative temperature-dependent changes in its fall time (purple symbols in Figure 3h) as well as its exponential recovery constant (blue symbols). This stands in stark contrast to the temperature-independent temporal response of the fast time-zero component, as evident when comparing individual time-delay traces in Figures 3e and 3f.

The absence of the slow dynamic component in the tr-MOKE signal at room temperature (see Figure 3b), followed by its emergence between 200-250 K and subsequent rapid increase below $T_C = 80$ K (see Figure 3g), confirms its magnetic origin. Simultaneously, qualitative temperature-dependent changes in the time constants (fall time and exponential recovery) suggest the transfer of spin angular momentum to the lattice (and, subsequently, back to the spin subsystem). This is in agreement with several recent works that have reported intriguing observations of the transfer of angular momentum between spin and lattice systems, and vice versa [69-72]. Specifically, at higher temperatures, the increased phonon population enhances the coupling between spin and lattice degrees of freedom, making the magnetoelastic interaction more efficient and, thus, faster. Conversely, at lower temperatures, reduced phonon availability limits interactions between spin and lattice, slowing down magnetoelastic coupling and affecting both the onset and recovery rates. Thus, the observed temperature-dependent variations in time constants (see Figure 3h) can be explained by considering the influence of phonon density on the efficiency of spin-phonon interactions.

In summary, our time- and temperature-dependent THz-pump tr-MOKE measurements helped identify and disentangle several different ultrafast responses of the quasi-2D interfacial ferromagnetic state at the LaNiO$_3$/CaMnO$_3$ interface to an intense single-cycle THz electric-field pulse – a sub-ps time-zero response consistent with ultrafast demagnetization driven by the nonequilibrium THz-induced spin-polarized currents, and multi-ps-scale change in the magnetic state of the superlattice due to the transfer of spin angular momentum to the lattice. These magnetic and magnetoelastic responses are observed concomitantly with the purely-electronic effects, such as the non-equilibrium Joule heating and a sudden partial metallization of the superlattice via direct interband tunnelling. These electronic and magnetic phenomena appear to be strongly interconnected, which suggests a new approach for controlling interfacial magnetism via ultrafast THz E-field pulses.

In order to gain better insight into the electronic dynamics initiated by the THz pulse, we have carried out ultrafast temperature-dependent THz-pump - near-IR (800 nm) reflectivity measurements of the 8LNO/4CMO superlattice in the same experimental setup. Due to the strong correlation between the THz-induced magnetic and electronic dynamics in this system, similar features (*fast* and *slow*) can also be observed in the electronic response of the superlattice measured via transient near-IR reflectivity (see Fig. S6 in the Supplementary Information). However, the temperature dependence of the amplitudes of the *fast*, *slow*, and quasiconstant dynamic ΔR/R responses does not align with the static SQUID magnetometry (M vs. T) measurements. Specifically, the ΔR/R vs. T curves (see Figure 4a) exhibit no change in slope at the $T_C$ (80 K). Instead, all three features exhibit a monotonic decline in intensity with rising temperature, as expected due to the temperature-dependent decline in conductivity (see electronic transport measurements in Fig. S3 of the Supplementary Information). Thus, these additional measurements suggest that the combination of ultrafast ΔR/R and tr-MOKE measurements can help decouple the electronic and magnetic responses of an interfacial ferromagnetic state.

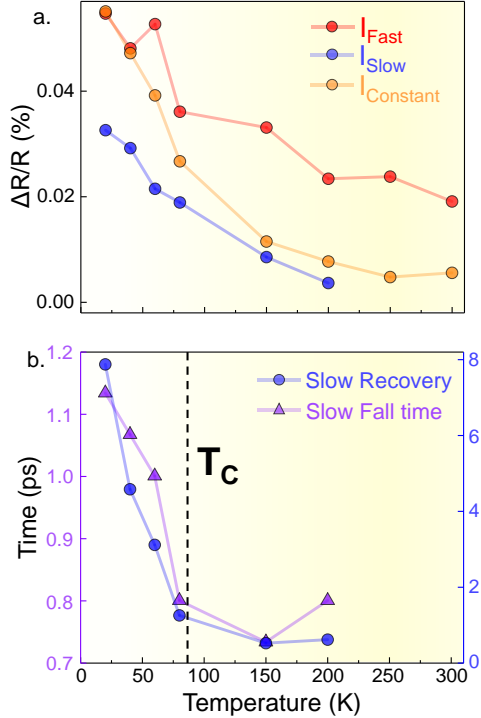

**Figure 4 | Temperature dependent ultrafast reflectivity and the electronic response of the superlattice. a.** Temporally decomposed amplitudes of the *fast*, *slow*, and quasiconstant dynamic components observed in the ultrafast ΔR/R response plotted as function of temperature. A monotonic decline in the intensities of all the components is observed with rising temperature. **b.** The temperature dependence of the ultrafast ΔR/R time constants of the *slow* feature exhibits signatures of the magnetic response of the system due to the strong correlation between the charge and spin dynamics in this system.

It is important to point out that the signatures of the magnetic response of the system persist in the temperature dependence of the ΔR/R *time constants* (fall time and the exponential recovery constant) of the *slow* feature due to its magnetic origin (see Fig. 4b). Such a correlation between the electronic and magnetic responses of the system offers a way of probing magnetic dynamics of the interfacial ferromagnetic state via ultrafast THz-pump near-IR reflectivity probe techniques.

The future undoubtedly holds many more exciting developments in the field of ultrafast THz spintronics, with the imminent advances in the ultrafast element- and spin-resolved probes of THz-driven dynamics, such as tr-ARPES [73] and x-ray free-electron laser-based techniques [74]. These novel probes of ultrafast electronic and magnetic dynamics will be vital in revealing the full time-, spin-, and momentum-resolved picture of the fundamental interaction at the interfaces of strongly correlated and quantum materials.

## METHODS

**XAS and XMCD:** The XAS and XMCD measurements were conducted using the Vector Magnet endstation at the high-resolution (100 meV) Magnetic Spectroscopy and Scattering beamline 4.0.2 at the Advanced Light Source [52]. The measurements were performed in the bulk-sensitive luminescence yield (LY) detection mode at a temperature of T = 20 K and with an in-plane magnetic field of 0.5 T. The X-ray beam (100 μm diameter) was incident on the sample at an angle of 30 degrees, as measured from the sample plane. Multiple measurements were carried out at various locations on the sample to rule out the possibility of X-ray beam damage.

**X-ray magnetic reflectivity (XRMR):** The magneto-optical profiling of the superlattice via XRMR was carried out using the Resonant X-ray Scattering end station at the same beamline. The samples were aligned with their surface-normal in the scattering plane and measured at T = 20 K and in an in-plane magnetic field of 0.1 T. The measurements were carried out in the specular θ–2θ reflection geometry at both resonant (Mn $L_3$ edge) and non-resonant (620 eV) photon energies using circularly polarized X-rays. The data were fitted using the X-ray reflectivity analysis program ReMagX [54] that uses an algorithm based on the Parratt formalism [75] and the Névot–Croce interdiffusion approximation [57]. For the non-resonant spectrum fitting, only the thicknesses of the $CaMnO_3$ and $LaNiO_3$ layers and the interdiffusion lengths between them were allowed to vary. For the resonant spectrum fitting, the variables included the thickness and roughness of the interfacial magnetic layer and the X-ray optical constant Δβm, which quantifies the magnitude of the modulation of the magnetic dichroism of the X-ray absorption coefficient β. The resonant X-ray optical constants required for the calculations were obtained by performing a Kramer-Kronig analysis of the XAS data and later optimized during the fitting of the resonant reflectivity spectra.

**Ultrafast THz measurements:** The ultrafast THz-pump near-IR probe measurements were conducted at Stockholm University's THz laboratory. The experiment utilized intense single-cycle THz electric fields, generated through optical rectification of 1300 nm pulses in an OH1 organic crystal. The generated THz field was modulated using a mechanical chopper at 500 Hz (half the laser repetition rate, 1 kHz), allowing for pump-on and pump-off measurements. In all experiments detailed in the main text, *s*-polarized THz pump pulses and *p*-polarized 800 nm probe pulses with a nominal pulse duration of 40 fs were employed. The measurements used a balanced detection scheme for tr-MOKE measurements, where photodiodes were balanced using a half-wave plate. For ultrafast reflectivity and transmittivity measurements, we used an unbalanced differential detection without any waveplates. All measurements were carried out within a cryostat, affording the flexibility to adjust temperatures within the range of 4 K to 300 K, enabling temperature-dependent investigations. A custom-made sample holder, equipped with permanent magnets of 0.47 T field strength, was used inside the cryostat to secure the sample. In the fluence-dependent measurements mentioned in Figure S6, two wire-grid polarizers were employed to control the THz pump fluence. The THz pump field was characterized through free space electro-optic sampling in a 50 μm-thick GaP crystal.

**HAXPES:** The bulk-sensitive core-level and valence-band HAXPES measurements were conducted at the P22 beamline [76] of the PETRA III synchrotron at DESY. The photon energy was set at 6.0 keV. At this energy, the values of the inelastic mean-free paths of the photoelectrons in $CaMnO_3$ and $LaNiO_3$ are estimated to be approximately 87 Å and 71 Å, respectively, with the maximum probing depth being roughly three times these values [77]. The total experimental energy resolution (380 meV at the analyzer pass energy of 50 eV) and the position of the zero binding energy were determined by measuring the Fermi edge of a standard Au sample. The measurements were carried out at the sample temperature of approximately 77 K.

**ACKNOWLEDGEMENTS**

A.M.D., J.R.P., and A.X.G. acknowledge support from the US Department of Energy, Office of Science, Office of Basic Energy Sciences, Materials Sciences and Engineering Division under awards number DE-SC0019297 and DE-SC0024132. A.X.G. also gratefully acknowledges the support from the Alexander von Humboldt Foundation. M.K., M.T., T.-C.W., and J.C. acknowledge the support by the U.S. Department of Energy, Office of Science, Office of Basic Energy Sciences under award number DE-SC0022160. This research used resources of the Advanced Light Source, which is a DOE Office of Science User Facility under Contract No. DE-AC02-05CH11231.

**AUTHOR CONTRIBUTIONS**
A.M.D. carried out the X-ray spectroscopic and scattering experiments under the supervision of A.X.G. and in collaboration with C.K. and P.S.  M.B. and V.U. carried out the ultrafast experiments under the supervision of S.B. and in collaboration with A.M.D. and A.X.G.  A.M.D. analyzed the data in collaboration with J.R.P., M.B., and V.U. and under the supervision of S.B. and A.X.G.  M.K., M.T., and T.-C.W. synthesized the samples and carried out structural and electronic transport characterization under the supervision of J.C.  J.P. carried out the HAXPES experiments under the supervision of A.X.G. and in collaboration with A.G. and C.S.  A.A. carried out magnetic characterization under the supervision of C.M.S. and in collaboration with A.X.G. All the co-authors participated in writing and editing the paper.

# Ultrafast terahertz field control of the emergent magnetic and electronic interactions at oxide interfaces

A. M. Derrico[1,2], M. Basini[3], V. Unikandanunni[3], J. R. Paudel[1], M. Kareev[4], M. Terilli[4], T.-C. Wu[4], A. Alostaz[5], C. Klewe[6], P. Shafer[6], A. Gloskovskii[7], C. Schlueter[7], C. M. Schneider[5], J. Chakhalian[4], S. Bonetti[3,8,*], and A. X. Gray[1,*]

[1] *Department of Physics, Temple University, Philadelphia, Pennsylvania 19122, USA*

[2] *Department of Physics, University of California, Berkeley, California 94720, USA*

[3] *Department of Physics, Stockholm University, 10691 Stockholm, Sweden*

[4] *Department of Physics and Astronomy, Rutgers University, Piscataway, New Jersey 08854, USA*

[5] *Peter Grünberg Institut (PGI-6), Forschungszentrum Jülich GmbH, D-52425 Jülich, Germany*

[6] *Advanced Light Source, Lawrence Berkeley National Laboratory, Berkeley, California 94720, USA*

[7] *Deutsches Elektronen-Synchrotron, DESY, 22607 Hamburg, Germany*

[8] *Department of Molecular Sciences and Nanosystems, Ca' Foscari University of Venice, 30172 Venice, Italy*

*email: stefano.bonetti@unive.it, axgray@temple.edu

**Structural Characterization via X-ray Diffraction (XRD) and Soft X-ray Reflectivity (SXR)**

Supplementary Figures S1(a) and S1(b) illustrate the high-resolution θ-2θ diffraction spectra of the 2LNO/4CMO and 8LNO/4CMO superlattices, respectively. The LaAlO$_3$ substrate peak appears at 2θ = 48º for both samples, in line with previous studies [1-3]. The 0$^{th}$-order superlattice peak (SL$_0$) for the thinner sample (2LNO/4CMO) is obscured by the substrate peak [Fig. S1(a)], aligning with earlier investigations of analogous LNO/CMO superlattices [1,3]. In the case of the thicker superlattice (8LNO/4CMO) [Fig. S1(b)], the same peak is observed at a slightly lower angle (47º), consistent with the aforementioned studies [1,3]. The two 1$^{st}$-order superlattice peaks (SL$_{-1}$ and SL$_{+1}$) for the thinner sample (2LNO/4CMO) are observed at symmetric angular positions of 44º and 52º, approximately ±4º from the 0$^{th}$-order (SL$_0$) peak. Conversely, for the 8LNO/4CMO superlattice, the two 1$^{st}$-order superlattice peaks (SL$_{-1}$ and SL$_{+1}$) are closer to the 0$^{th}$-order (SL$_0$) peak (±2.15º), as expected from basic considerations [1,4,5].

All superlattice peaks exhibit shapes characteristic of high-quality single-crystalline superlattices. Additionally, the spectrum for the thinner superlattice (2LNO/4CMO) displays pronounced SL thickness fringes. The number of aforementioned fringes (8) corresponds to the expected number (P − 2 = 8) of superlattice periods (P = 10). In the case of the thicker superlattice (8LNO/4CMO), the SL thickness fringes appear much closer and merge together, in line with prior studies [3,5]. A more comprehensive structural characterization of the thicker superlattice (8LNO/4CMO), central to this study, was conducted using synchrotron-based resonant and non-resonant soft X-ray reflectivity (SXR), and the details are described below.

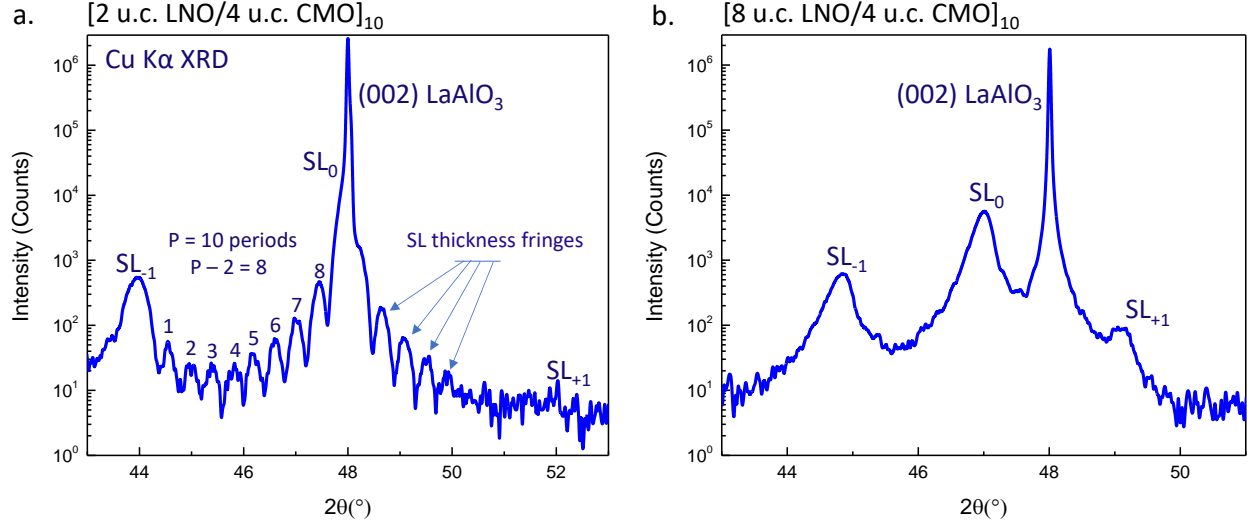

**Figure S 1. a.** θ-2θ X-ray diffraction spectrum of the 2LNO/4CMO superlattice. The $0^{th}$-order superlattice peak ($SL_0$) is obscured by the (002) $LaAlO_3$ substrate peak at 2θ = 48°. The anticipated number of SL thickness fringes is eight (P − 2 = 8) for a superlattice with ten periods (P = 10). **b.** θ-2θ spectrum for the 8LNO/4CMO superlattice. The $0^{th}$-order superlattice peak ($SL_0$) is detected at 2θ = 47°. Both superlattices exhibit prominent $1^{st}$-order superlattice peaks labeled as $SL_{-1}$ and $SL_{+1}$.

Supplementary Figures S2(a) and S2(b) display $q_z$-dependent specular soft X-ray reflectivity (SXR) spectra recorded using photon energies corresponding to off-resonant (620 eV) and resonant (Mn $L_3$ at 642.7 eV) conditions, respectively. The spectra cover a broad range of $q_z$ (0.05 - 0.35 1/Å), encompassing both the $1^{st}$-order and $2^{nd}$-order Bragg conditions (at ~0.15 and ~0.30 1/Å, respectively), and thus provide detailed depth-resolved information on both the layering and interfacial structure of the sample [6-8]. The experimental data (red curves) were self-consistently fitted with the SXR analysis program ReMagX [7], employing an algorithm based on the Parratt formalism [9] and the Névot–Croce interdiffusion approximation [10]. In the model, only the thicknesses of the $CaMnO_3$ and $LaNiO_3$ layers, and the interdiffusion lengths between them (interface roughness), were allowed to vary. The resonant X-ray optical constants required for fitting the on-resonance data were obtained through a Kramer-Kronig analysis of the XAS data in Figure 1a of the main text.

The blue-colored spectra depicted in Supplementary Figures S2(a-b) illustrate optimal theoretical fits to the experimental data, while Figures S2(c-d) show the resulting X-ray optical profile of the 8LNO/4CMO superlattice. The aforementioned superlattice profile is represented as the depth-dependent variation of the X-ray absorption coefficient $\beta$ at the photon energy corresponding to the Mn $L_3$ absorption edge. The maxima in such an element-selective (Mn) absorption profile correspond to the depth-resolved positions of the $CaMnO_3$ layers, and the minima to the positions of the $LaNiO_3$ layers, where the Mn element is absent. The individual layer thicknesses obtained from the X-ray optical fitting are 28.46 ± 0.22 Å for $LaNiO_3$ and 14.25 ± 0.08 Å for $CaMnO_3$. These values correspond to approximately 7.5 u.c. of $LaNiO_3$ and approximately 4 u.c. of $CaMnO_3$, using lattice constants from prior studies [2-4,11,12]. The average interface roughness (interdiffusion) is 3.93 ± 0.54 Å, which equates to approximately one unit cell of a typical perovskite oxide and aligns with typical high-quality layer-by-layer growth [13]. This observed interface roughness may account for the slightly underestimated (by 0.5 u.c.) average thickness of the $LaNiO_3$ layer (nominally 8 u.c.).

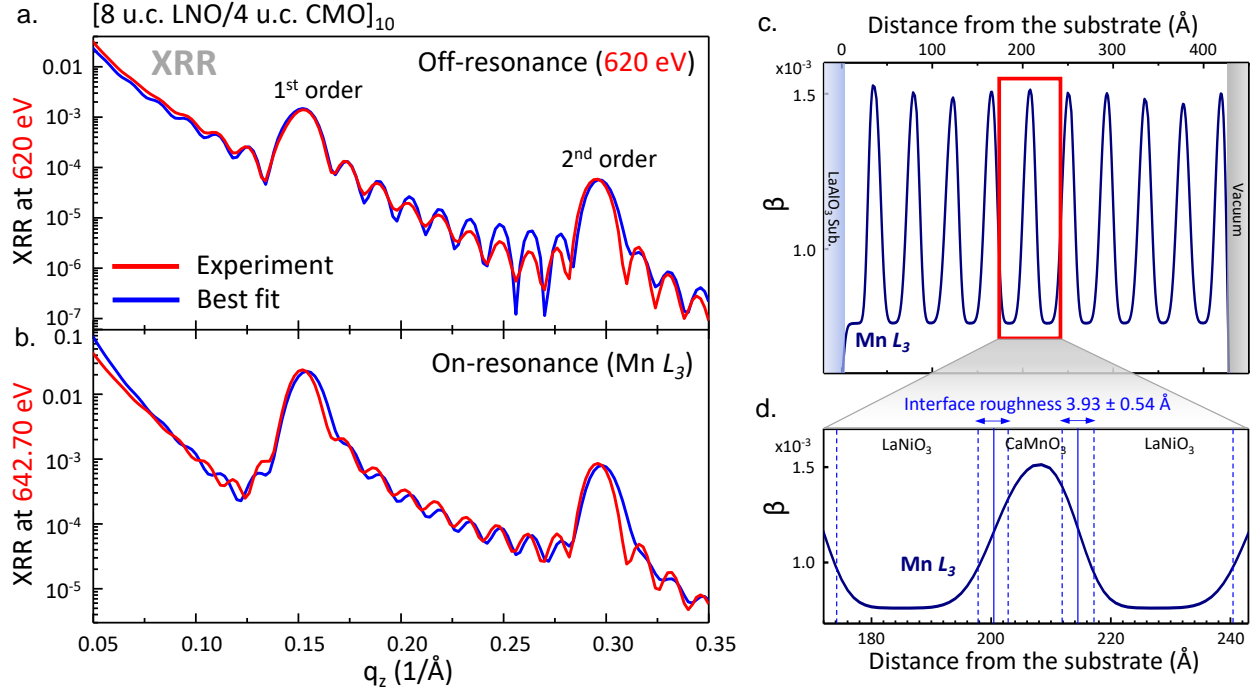

**Figure S 2. a.** and **b.** Momentum-dependent SXR spectra (red curves) and the best fits (blue curves) to the experimental data measured at the photon energies corresponding to off-resonant (620 eV) and resonant (Mn $L_3$ at 642.7 eV) conditions, respectively. Self-consistent fitting of the data yields a detailed optical (absorption coefficient β) profile of the sample, shown in **c.** and **d.**, with the resultant layer thicknesses of 28.46 ± 0.22 Å (~7.5 u.c. of LaNiO$_3$) and 14.25 ± 0.08 Å (~4 u.c. CaMnO$_3$), as well as the average interface roughness (chemical interdiffusion) of 3.93 ± 0.54 Å.

### Chemical, Electronic, and Magnetic Characterization

The nominal chemical composition of the superlattices was confirmed through bulk-sensitive HAXPES measurements conducted at the P22 beamline [14] of the PETRA III synchrotron at DESY. Supplementary Figure S3(a) presents wide-energy range HAXPES survey spectra for the 8LNO/4CMO (red line) and 2LNO/4CMO (blue line) superlattices. The presence of all expected elements (Ca, Mn, O, La, Ni, and C from the surface-adsorbed contaminant C/O layer) is verified by corresponding core-level peaks. The greater total thickness of the LaNiO$_3$ layers in the red 8LNO/4CMO spectrum is evidenced by the higher relative intensities of the La $3d$ and $4p$ peaks, compared to the blue 2LNO/4CMO spectrum.

To investigate the thickness-dependent variation in the valence-band electronic structure of LaNiO$_3$ and its impact on the resistivity of the superlattices, we employed a combination of bulk-sensitive valence-band HAXPES spectroscopy and electronic transport measurements. Supplementary Figure S3(b) displays the experimental valence-band spectra of the 8LNO/4CMO superlattice (red line) and the 2LNO/4CMO superlattice (blue line). The corresponding temperature-dependent sheet resistance curves, measured using the standard van der Pauw method, are shown in the inset.

The near-Fermi-level region of the red 8LNO/4CMO superlattice spectrum exhibits two prominent features at 0.3 eV and 1.0 eV which, based on prior studies, correspond to the strongly-hybridized Ni $3d$ $e_g$ and $t_{2g}$ states, respectively [3,15,16]. In line with earlier research on the thickness-dependent metal-insulator transition in LaNiO$_3$ [3,15,16], the superlattice containing below-critical-thickness LaNiO$_3$ layers (2LNO/4CMO) shows a significant suppression of these near-Fermi-level electronic states, resulting in an approximately two orders of magnitude enhancement in sheet resistivity.

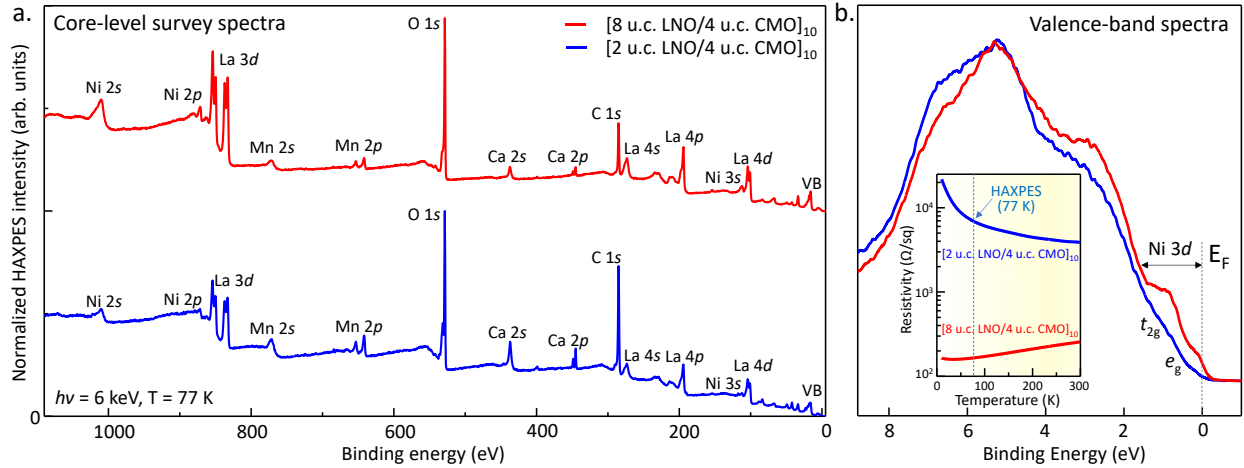

**Figure S 3. a.** Bulk-sensitive HAXPES survey spectra for the 8LNO/4CMO (red line) and 2LNO/4CMO (blue line) superlattices. The presence of all expected elements (Ca, Mn, O, La, Ni, and C from the surface-adsorbed contaminant C/O layer) is confirmed by corresponding core-level peaks. **b.** Angle-integrated valence-band spectra of the same superlattices recorded with a photon energy of 6 keV. Significant suppression of the near-$E_F$ Ni $3d$ $e_g$ and $t_{2g}$ density of states results in a metal-insulator transition in the 2LNO/4CMO sample, as probed with sheet resistivity measurements shown in the inset.

In the context of interfacial ferromagnetism, in the thicker 8LNO/4CMO superlattice, the aforementioned Ni $3d$ states facilitate charge transfer from Ni to the interfacial Mn sites, creating an electronic environment that stabilizes the ferromagnetic state mediated by the double exchange interaction [1-5,17]. The depletion of the Ni $3d$ $e_g$ states in the thinner 2LNO/4CMO superlattice leads to the metal-insulator transition in LaNiO$_3$ and the concomitant suppression of the interfacial ferromagnetic state in CaMnO$_3$.

Below, we probe the emergent interfacial ferromagnetic state in the 8LNO/4CMO superlattice using temperature-dependent SQUID magnetometry and compare our results with a prior study of a similar superlattice [5].

Supplementary Figure S4 shows the temperature dependence of the Mn magnetic moment from 30 to 300 K after field-cooling in 7 T with a 2 T warming field (red symbols, left $y$-scale). The magnetic moment was normalized to the estimated number of interfacial Mn ions in the superlattice. The observed saturated magnetic moment of approximately 1.1 $\mu_B$ per interfacial Mn at T = 20 K is in excellent agreement with the previously reported value of ~1 $\mu_B$ per interfacial Mn [1]. A standard background subtraction procedure described in Ref. [18] was utilized to isolate the superlattice contribution to the magnetic signal and to account for the substrate (LaAlO$_3$) contribution.

We compare our data to the M vs. T curve obtained in a prior study [5] on a similar (6LNO/4CMO) superlattice (blue curve, right $y$-axis). This data exhibits a slightly better signal-to-noise ratio due to the lower value (0.5 T) of the warming field used during the measurement. This allows for the observation of the finer details of the temperature dependence of the magnetic moment, such as the slow onset of the ferromagnetic signal between 250 and 200 K, as well as the abrupt change of the slope at $T_C = 80$ K. In summary, the two datasets exhibit good qualitative agreement. This agreement becomes quantitative when we also consider the observed values of the saturated magnetic moment at high fields (1.1 $\mu_B$/Mn vs. 1 $\mu_B$/Mn) [1].

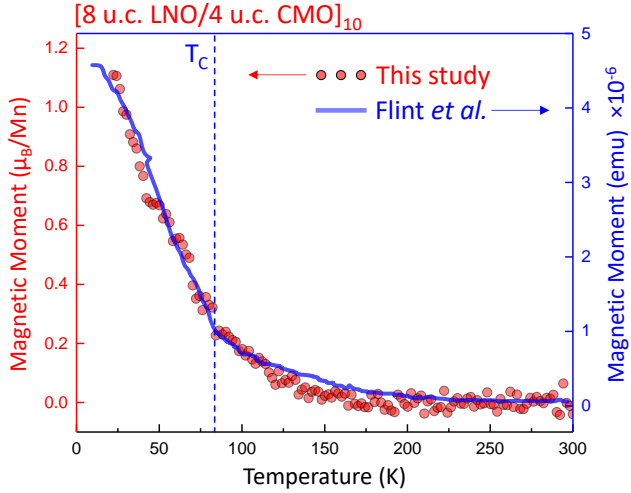

**Figure S 4.** Temperature dependence of the Mn magnetic moment for the 8LNO/2CMO superlattice obtained in a warming field of 2 T after field-cooling in a 7 T field (red symbols, left y-scale). The standard background subtraction procedure described in Ref. [18] was utilized to isolate the superlattice contribution to the magnetic signal. Subsequently, the data were normalized to the estimated number of interfacial Mn ions in the superlattice to obtain the values of $\mu_B$ per interfacial Mn. The data are compared to the M vs. T curve obtained in a prior study [5] on a similar (6LNO/4CMO) superlattice (blue curve, right y-axis).

**Fluence-Dependent THz-pump tr-MOKE Measurements (8LNO/4CMO sample, T = 20 K)**

Supplementary Figure S5 below presents a series of fluence-dependent THz-pump tr-MOKE time-delay traces recorded on the 8LNO/2CMO superlattice at T = 20 K. The THz-pump excitation strengths are expressed in terms of the peak E-field of the THz pulse, measured and calibrated using standard electro-optic sampling. The fluence (peak E-field) was varied from the maximum value of 1.03 MV/cm to the minimum value of 0.12 MV/cm using a pair of rotatable THz wire-grid polarizers. The individual time-delay traces recorded at several values of the THz peak E-field within this range are displayed in panel (a). Panel (b) presents the average amplitudes of the magnetic response (Kerr rotation angle $\phi_K$) for the 'fast' time-zero response (peak), obtained by averaging over several time delays around the 'fall time' of approximately +250 fs. The solid black curve represents the best fit of the experimental data to a parabola.

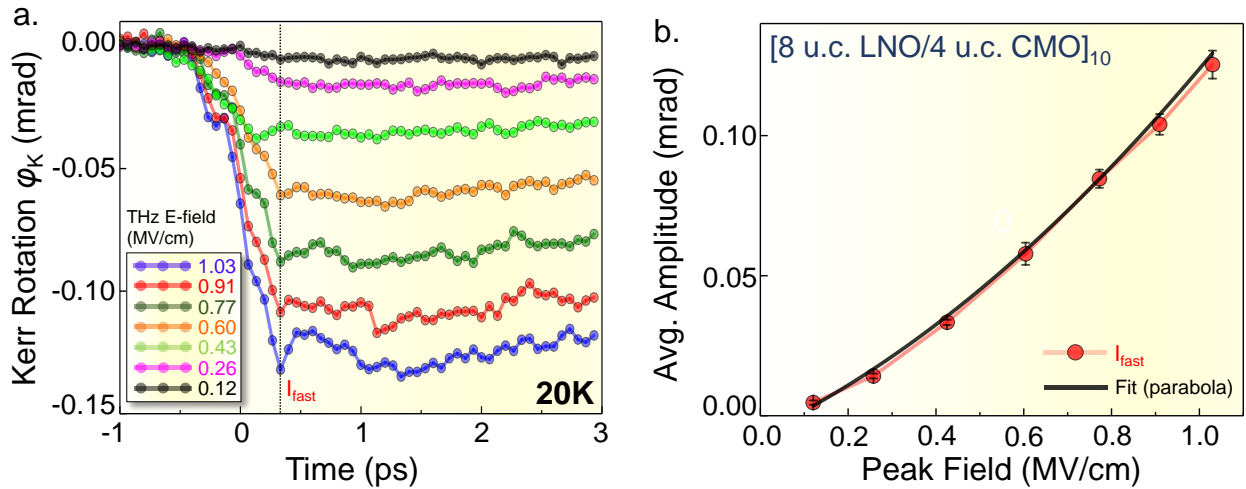

**Figure S 5. a.** Fluence (THz peak E-field) dependent THz-pump tr-MOKE time-delay traces recorded on the 8LNO/2CMO superlattice at T = 20 K. **b.** Average amplitudes of the magnetic response (Kerr rotation angle $\phi_K$) for the 'fast' electronic response (red symbols) and the best fit of the experimental data to a parabola (black solid line).

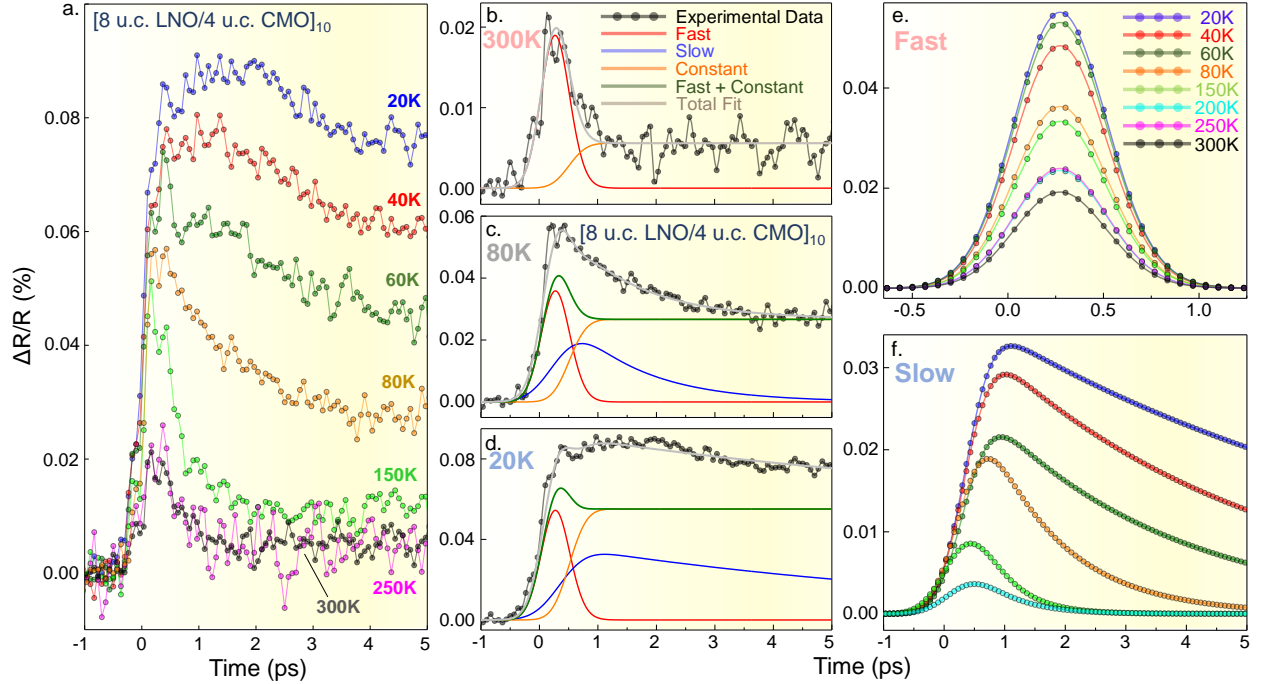

**Figure S 6. a.** Temperature-dependent ΔR/R delay traces recorded at several temperatures varying from 300 to 20 K. **b.** Temporal decomposition of the room-temperature (300 K) delay trace into the initial fast and quasiconstant components. **c.** Temporal decomposition of the delay trace recorded at $T_C$ = 80 K requires an additional dynamic component characterized by a slower risetime and consequent exponential recovery (blue curve). **d.** Temporal decomposition of the 20 K delay trace requires all three dynamic components (fast, quasiconstant, and slow). **e.** All the temperature-dependent fits of the fast dynamic component, exhibiting uniform temporal behavior. **f.** Temperature-dependent fits of the slow dynamic component, which becomes prominent at the onset of interfacial ferromagnetism (T < 80 K).

## Ultrafast Temperature-Dependent THz-pump near-IR Reflectivity (ΔR/R) Measurements

Supplementary Figure S6 displays the results of additional ultrafast temperature-dependent THz-pump near-IR (800 nm) reflectivity probe measurements of the 8LNO/4CMO superlattice in the same experimental setup. Panel (a) shows the time-resolved electronic response of the sample measured via relative (%) change in transient IR reflectivity (ΔR/R) at several temperatures varying from 300 K (room temperature) to 20 K. Panels (b)-(d) illustrate temporal decomposition of the delay traces recorded at room temperature (b), $T_C$ = 80 K (c), and T = 20 K (d) into three distinct dynamic components (fast, quasiconstant, and slow) using the fitting procedure described in the main text. Panels (e) and (f) show all the resultant temperature-dependent fits of the 'fast' and 'slow' dynamic temporal components, respectively. Results of the analyses of the amplitudes and time constants are shown (and discussed) in Figure 4 of the main text.